\documentstyle[epsfig]{mn2e}

\title[SDSS\,J0349-0059 is a GW Virginis star]
{SDSS\,J0349-0059 is a GW Virginis star}
\author[Patrick A.~Woudt, Brian Warner and Ewald Zietsman]
       {Patrick A.~Woudt$^1$\thanks{email: Patrick.Woudt@uct.ac.za},
Brian Warner$^{1,2}$\thanks{email: Brian.Warner@uct.ac.za} and Ewald Zietsman$^{1}$\\
        $^1$Astrophysics, Cosmology and Gravity Centre, Department of Astronomy, 
        University of Cape Town, Private Bag X3, \\ 
        Rondebosch 7701, South Africa\\
        $^2$School of Physics and Astronomy, Southampton University, Southampton, UK}
\date{Accepted 2012 August 8.  Received 2012 August 8; in original form 2012 July 31}

\begin{document}

\maketitle

\begin{abstract}
High speed photometric observations of the spectroscopically-discovered 
PG 1159 star SDSS\,J034917.41-005917.9 in 2007 and 2009
reveal a suite of pulsation frequencies in the range of 1038 -- 3323 $\mu$Hz with amplitudes
between 3.5 and 18.6 mmag. SDSS\,J034917.41-005917.9 is therefore a member of the GW Vir class
of pulsating pre-white dwarfs. We have identified 10 independent pulsation frequencies that can be fitted by an
asymptotic model with a contant period spacing of 23.61 $\pm$ 0.21 s, presumably associated with a 
sequence of $\ell = 1$ modes. The highest amplitude peak in the suite of frequencies 
shows evidence for a triplet structure, with a frequency separation of 14.4 $\mu$Hz. Five of the
identified frequencies do not fit the $\ell = 1$ sequence, but are, however, well-modeled by an 
independent asymptotic sequence with a constant period spacing of 11.66 $\pm$ 0.13 s. 
It is unclear to which $\ell$ mode these frequencies belong.

\end{abstract}


\section{Introduction}

The GW Vir stars are a sub-group of variables in the spectroscopic PG 1159 class, 
which form a link between the (post-AGB) central stars of planetary 
nebulae and the H-deficient white dwarf cooling sequence. They pulsate 
non-radially and lie in an instability strip bounded by effective 
temperatures $200\,000 > T_{\rm eff} > 75\,000$ K, excited by the kappa
mechanism working through partial ionization of carbon and oxygen. 
Studying these stars with astroseismology has provided important knowledge 
on the interiors of the late stages of stellar evolution (Winget \& Kepler 2008). 
There are 19 known GW Vir stars (Quirion, Fontaine \& Brassard 2007; Quirion 2009), showing a wide 
variety of behaviour. The possible addition of more examples is therefore 
of significance. Here we show that the known spectroscopic PG 1159 star 
SDSS J034917.41-005919.2 (hereafter SDSS J0349-0059) is a non-radial 
pulsator, putting it in the GW Vir subclass. In Sect.~2 we describe 
what is already known about SDSS J0349-0059 and list our high speed 
photometric observations. Sect.~3 analyses these and presents 
comparisons with other GW Vir stars.

\section{SDSS\,J0349-0059 as a PG 1159 star}

\subsection{The PG 1159 Instability Strip}

H\"ugelmeyer et al.~(2006) used Sloan Digital Sky Survey spectra 
of five spectroscopically discovered PG 1159 stars, including SDSS\,J0349-0059, to interpret their 
spectra using non-LTE model atmospheres; these were added to the analyses 
of six previously analysed stars. Among these, SDSS\,J0349-0059 has an 
average position, with $T_{\rm eff}$ = 90.0 $\pm$ 0.9 kK and 
$\log g = 7.50 \pm 0.01$ (cgs).

C\'orsico et al.~(2006) computed nonadiabatic pulsation models 
for stars in the GW Vir instability strip and found that their models 
agree with the observed strip. The parameters listed above 
for SDSS J0349-0059 place it within the observed instability strip 
(cf. figures 6 and 7 of C\'orsico et al.).

\subsection{High speed photometry of SDSS\,J0349-0059}

SDSS~J0349-0059 was observed in January 2007, March 2009 and December 
2009 with the University of Cape Town's CCD photometer (O'Donoghue 1995) 
attached to the 40-in and 74-in reflectors at the Sutherland site of the South 
African Astronomical Observatory. The motivation for observing this star 
was inclusion as a UV-rich star included in a search for possible AM CVn stars.

\begin{figure}
\centerline{\hbox{\psfig{figure=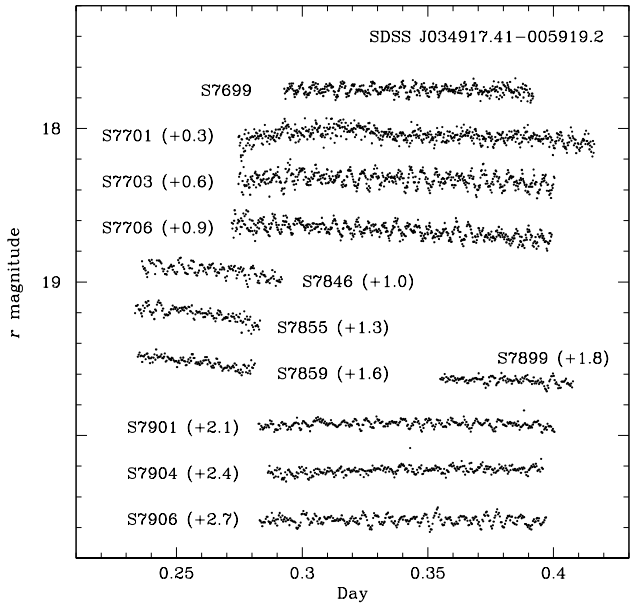,width=8.4cm}}}
  \caption{Individual light curves of SDSS\,J0349-0059. The light curve
of run S7699 is displayed at the correct brightness. Vertical offsets (indicated 
between brackets) have been applied to other light curves for display purposes
only.}
 \label{sdss0349fig1}
\end{figure}

The observing log is given in Tab.~\ref{sdss0349tab1} and the individual light curves 
are shown in Fig.~\ref{sdss0349fig1}. Note that the short observing runs in March 2009
(S7846, S7855 and S7859) were observed at high airmass ($1.7-3$); 
this is evident in the downwards trend seen in each of the light curves.
The differential photometric correction fails at these large airmasses 
due to the different 
colours of the target and the reference star; SDSS\,J0349-0059 has $(g-r) = -0.42$, where the reference
star is redder at $(g-r)= 0.26$ (SDSS Data Release 8: Aihara et al.~2011).

\begin{figure}
\centerline{\hbox{\psfig{figure=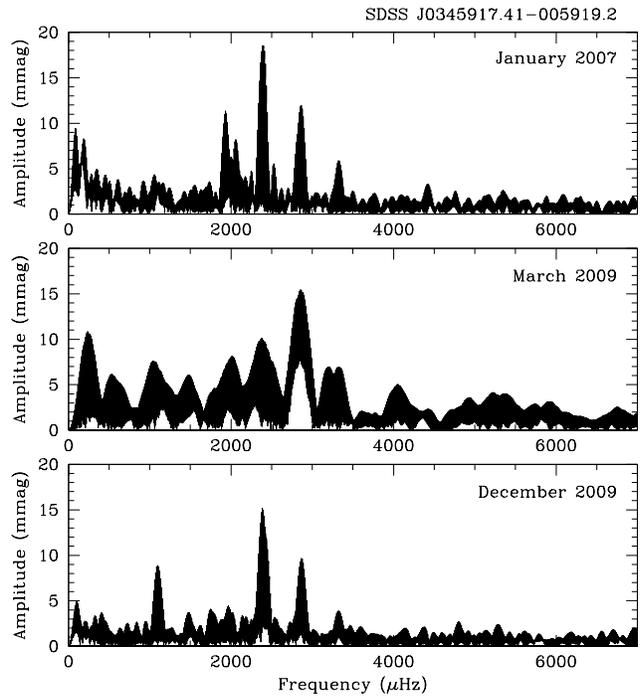,width=8.4cm}}}
  \caption{The Fourier transforms of SDSS\,J0349-0059 in January 2007
(upper panel), March 2009 (middle panel) and December 2009 (lower panel).}
 \label{sdss0349fig2}
\end{figure}

The Fourier transforms (FTs) of the three
data sets (January 2007, March 2009 and December 2009) are shown in
Fig.~\ref{sdss0349fig2}. In the construction of the FTs, individual runs have their
mean and trend removed.

Rapid oscillations with a range $\sim 0.2$ mag and
time scale $\sim 300-900$ s are visible in the light curves and their FTs, 
which, with the spectra discussed 
by H\"ugelmeyer et al.~(2006), place SDSS J0349-0059 in the GW Vir 
subgroup of PG 1159 stars.

\begin{table}
 \centering
  \caption{Observing log of photometric observations}
   \begin{tabular}{@{}llccrl@{}}
 Run      & Date of obs.          & HJD$^\dag$ & Length  & t$_{in}$  &  V \\
          & (start of night)      &  (+2450000.0)     & (h)     & (s)      & (mag) \\[10pt]
 S7699    & 20/01/2007   & 4121.29287 & 2.38 &  20  & 17.7 \\
 S7701    & 21/01/2007   & 4122.27472 & 3.39 &  20  & 17.7 \\
 S7703    & 22/01/2007   & 4123.27466 & 3.02 &  20  & 17.7 \\
 S7706    & 23/01/2007   & 4124.27198 & 3.06 &  20  & 17.8 \\
 S7846    & 20/03/2009   & 4911.23629 & 1.34 &  30  & 17.9 \\
 S7855    & 23/03/2009   & 4914.23355 & 1.19 &  30  & 17.9 \\
 S7859    & 24/03/2009   & 4915.23460 & 1.12 &  30  & 17.9 \\
 S7899    & 24/12/2009   & 5190.35493 & 1.27 &  30  & 17.8 \\
 S7901    & 25/12/2009   & 5191.28264 & 2.83 &  30  & 17.8 \\
 S7904    & 26/12/2009   & 5192.28631 & 2.63 &  30  & 17.8 \\
 S7906    & 27/12/2009   & 5193.28308 & 2.74 &  30  & 17.9 \\[5pt]
\end{tabular}
{\footnotesize
\newline
$^\dag$HJD of first observation; t$_{in}$ is the integration time. \hfill}
\label{sdss0349tab1}
\end{table}
   
\section{Analysis}

The Fourier transforms of the combined runs in January 2007 (upper panel of 
Fig.~\ref{sdss0349fig2}) and December 2009 (lower panel of Fig.~\ref{sdss0349fig2})
show a substantial amount of variation in the distribution of power in the FTs. The March 2009
data set is shown for completeness, but given the relatively few short observing runs, the
signal-to-noise in the FT of the combined runs in March 2009 (middle panel of
Fig.~\ref{sdss0349fig2}) is too low to identify many individual pulsation frequencies.
However, the main peak in March 2009 at $\sim 2865$ $\mu$Hz is also seen in 
January 2007 and December 2009.

The distribution of pulsation frequencies in SDSS J0349-0059 resembles closely that of the 
GW Vir star PG\,1707+427 (Kawaler et al.~2004), which has a main peak at $2236$ $\mu$Hz and 
additional frequencies in the range of 1100 -- 2988 $\mu$Hz. In SDSS\,J0349-0059 the main peak is located
at 2386 $\mu$Hz with additional pulsation frequencies in the range of 1097 -- 3323 $\mu$Hz.

\begin{figure*}
\centerline{\hbox{\psfig{figure=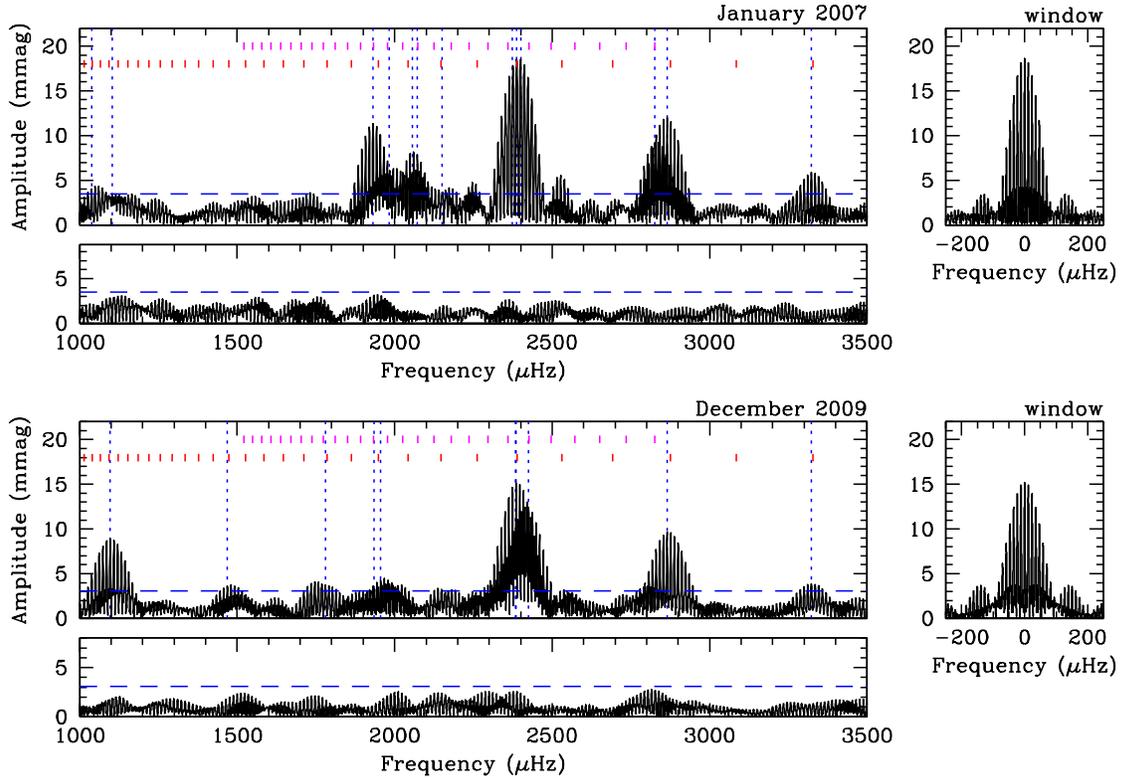, width=16cm}}}
  \caption{The Fourier transforms of SDSS\,J0349-0059 in January 2007
(upper panels) and December 2009 (lower panels). For each data set three
panels are displayed: the original Fourier transform of the combined observing
run (top-left), the residual Fourier transform after prewhitening with the
identified frequencies listed in Tab.~\ref{sdss0349tab2} (lower-left) and
the window function (top right). The $4\sigma$ amplitude limit is indicated by the
horizontal dashed line, the identified frequencies are marked by the vertical dashed
lines and frequencies associated with the two asymptotic models with a constant period 
spacing are marked by short vertical markers, see Sect.~3.2 for details.}
 \label{sdss0349fig3}
\end{figure*}

Fig.~\ref{sdss0349fig3} shows an expanded view of the FT of the combined runs in January 2007 
(upper two panels) and December 2009 (lower two panels) in the 1000 -- 3500 $\mu$Hz frequency range.
The windows functions for the two data sets are given in the rightmost panels of 
Fig.~\ref{sdss0349fig3}, scaled to the amplitude of the highest peak in the respective FTs.
In each FT, the horizontal dashed line corresponds to $4\sigma$ amplitude
detection limit; this limit is 3.5 mmag for January 2007 and 3.1 mmag for December 2009. 
The standard deviation ($\sigma$) has been determined over the 1000 -- 4000 $\mu$Hz frequency
range, after prewhitenening the FT by the frequencies listed in Tab.~\ref{sdss0349tab2}
and randomizing the residual brightness variations for each observing run 
against the timing array.

In the January 2007 FT, 13 frequencies have been identified with amplitudes above the $4\sigma$ 
amplitude limit. They are marked by the vertical dash lines in Fig.~\ref{sdss0349fig3}. 
Successive prewhitening of the highest amplitude peak has resulted in the 
identification of the 13 pulsation frequencies listed in Tab.~\ref{sdss0349tab2}. Some
of these frequencies could possibly be mis-identified as a one-day alias, although
the coincidence of frequencies in the two independent data sets gives some reassurance 
of correct identification; in Tab.~\ref{sdss0349tab2} we merely list the 
frequencies associated with the highest amplitude signals. The final
FT prewhitened at these 13 frequencies is shown immediately below the original FT.

Similarly, ten independent frequencies have been identified in the December 2009 data. They 
are listed in Tab.~\ref{sdss0349tab2} and marked by the vertical dashed lines in
the third vertical panel (left) of Fig.~\ref{sdss0349fig3}. Again, the
final FT prewhitened at the identified frequencies is displayed below the original FT.

The highest amplitude peak in both the 2007 and 2009 observations is located at
$\sim$ 2386 $\mu$Hz (peak amplitude of 18.6 and 15.2 mmag in 2007 and 2009, respectively). 
Other pulsation frequencies in common between the two data sets are around
2865 $\mu$Hz -- peak amplitudes of 12.1 mmag (2007) and 9.6 mmag (2009) -- and 3323 $\mu$Hz 
-- peak amplitudes of 5.5 mmag (2007) and 4.3 mmag (2009).

\begin{table}
 \centering
 \caption{Frequency identifications}
 \begin{tabular}{@{}cccc@{}}
Frequency & Period & Amplitude & Rank \\
($\mu$Hz) & (s)    & (mmag)    &      \\[10pt]
\multicolumn{4}{c}{January 2007}\\[2pt]
$1037.9 \pm 0.4$   & \hfill $963.48 \pm 0.37$  & \hfill $ 3.7 \pm 0.9$ & 12 \\
$1103.3 \pm 0.4$   & \hfill $906.37 \pm 0.33$  & \hfill $ 3.9 \pm 0.9$ & 11 \\
$1931.1 \pm 0.1$   & \hfill $517.84 \pm 0.03$  & \hfill $11.3 \pm 1.0$ & 4 \\
$1983.4 \pm 0.2$   & \hfill $504.18 \pm 0.05$  & \hfill $ 6.8 \pm 0.9$ & 6 \\
$2055.9 \pm 0.4$   & \hfill $486.40 \pm 0.09$  & \hfill $ 4.2 \pm 0.9$ & 10 \\
$2072.2 \pm 0.2$   & \hfill $482.58 \pm 0.05$  & \hfill $ 6.6 \pm 0.9$ & 8 \\
$2150.3 \pm 0.4$   & \hfill $465.05 \pm 0.09$  & \hfill $ 3.5 \pm 0.9$ & 13 \\
$2372.6 \pm 0.2$   & \hfill $421.48 \pm 0.04$  & \hfill $ 6.7 \pm 0.9$ & 7 \\
$2387.2 \pm 0.1$   & \hfill $418.90 \pm 0.02$  & \hfill $18.6 \pm 1.1$ & 1 \\
$2401.4 \pm 0.1$   & \hfill $416.42 \pm 0.02$  & \hfill $15.2 \pm 1.0$ & 2 \\
$2826.5 \pm 0.2$   & \hfill $353.79 \pm 0.03$  & \hfill $ 7.2 \pm 0.9$ & 5 \\
$2865.2 \pm 0.1$   & \hfill $349.02 \pm 0.01$  & \hfill $12.1 \pm 1.0$ & 3 \\
$3323.0 \pm 0.3$   & \hfill $300.93 \pm 0.03$  & \hfill $ 5.5 \pm 0.9$ & 9 \\[5pt]
\multicolumn{4}{c}{December 2009}\\[2pt]
$1097.1 \pm 0.1$   & \hfill $911.49 \pm 0.08$  & \hfill $ 9.1 \pm 0.7$ & 4 \\
$1468.8 \pm 0.3$   & \hfill $680.83 \pm 0.14$  & \hfill $ 3.6 \pm 0.7$ & 8 \\
$1779.9 \pm 0.3$   & \hfill $561.83 \pm 0.09$  & \hfill $ 4.0 \pm 0.7$ & 7 \\
$1935.3 \pm 0.4$   & \hfill $516.72 \pm 0.11$  & \hfill $ 3.3 \pm 0.7$ & 10 \\
$1955.3 \pm 0.3$   & \hfill $511.43 \pm 0.08$  & \hfill $ 4.2 \pm 0.7$ & 6 \\
$2383.5 \pm 0.4$   & \hfill $419.55 \pm 0.07$  & \hfill $ 3.4 \pm 0.7$ & 9 \\
$2385.6 \pm 0.1$   & \hfill $419.18 \pm 0.02$  & \hfill $15.2 \pm 0.9$ & 1 \\
$2425.6 \pm 0.2$   & \hfill $412.27 \pm 0.03$  & \hfill $10.0 \pm 0.8$ & 2 \\
$2864.8 \pm 0.1$   & \hfill $349.06 \pm 0.01$  & \hfill $ 9.6 \pm 0.8$ & 3 \\
$3323.0 \pm 0.3$   & \hfill $300.93 \pm 0.03$  & \hfill $ 4.3 \pm 0.7$ & 5 \\[5pt]
 \end{tabular}
  \label{sdss0349tab2}
\end{table}

\subsection{Rotational splitting of the 2387-$\mu$Hz (419 s) oscillation}

\begin{figure}
\centerline{\hbox{\psfig{figure=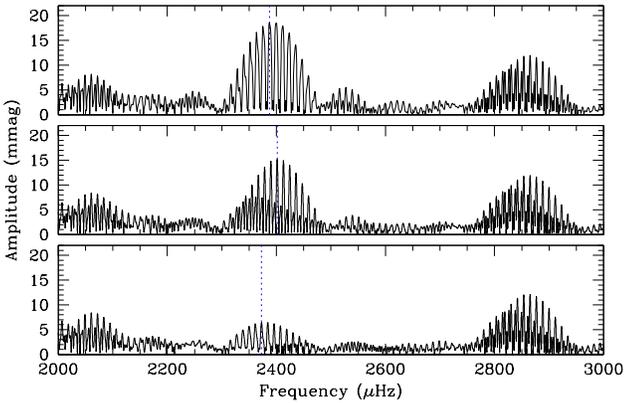,width=8.4cm}}}
 \caption{The Fourier transform of SDSS\,J0349-0059 in January 2007 (upper panel).
The middle panel displays the Fourier transform after prewhitening of the strongest
signal in the original data at 2387.2 $\mu$Hz. The lower panel displays 
the residual Fourier transform after subsequently prewhitening the data at a frequency of 
2401.4 $\mu$Hz. Peak frequencies of the resolved triplet at 419 s are marked by the
dashed vertical lines.}
 \label{sdss0349fig4}
\end{figure}

The highest peak in the 2007 data shows clear evidence for three closely spaced frequencies.
This is illustrated in Fig.~\ref{sdss0349fig4} where we show the January 2007 FT in the frequency
range of 2000 -- 3000 $\mu$Hz (original, top panel). After prewhitening by the main peak at 2387.2
$\mu$Hz, a clear window function remains (middle panel) at 2401.4 $\mu$Hz. Prewhitening at 2401.4 $\mu$Hz
leaves a distinct signal at 2372.6 $\mu$Hz (lower panel). All frequencies are marked by vertical dashed
lines. This suggests that the strongest peak is in fact a triplet, with a frequency separation of
14.4 $\pm$ 0.2 $\mu$Hz.

The 2009 data reveal two components of the triplet; a strong peak at 2385.6 $\mu$Hz and a low amplitude
peak at 2383.5 $\mu$Hz which is probably the 1-day alias of the 2372.6 $\mu$Hz peak seen in 2007.

We draw again on the similarity with PG\,1707+427, where evidence was found for a 9 $\mu$Hz split of the
main mode, interpreted as a rotational frequency splitting, corresponding to a rotation period of 0.65 
days (Kawaler {et al.}~2004). In SDSS\,J0349-0059 this would imply a rotation period of 0.40 $\pm$ 0.01 
days if the pulsation mode is $\ell = 1$.

\subsection{Period spacing}

Given the strong similarity with PG\,1707+427, we suspected that the frequencies
identified in SDSS\,J0349-0059 and listed in Tab.~\ref{sdss0349tab2} are predominantly part
of a sequence of $\ell = 1$ $g$-modes with constant period spacing; this is an asymptotic
model of the form $P_{model} = n \Delta P + C$, where $P_{model}$ is the model period, $n$ is an integer,
$\Delta P$ is a constant period spacing and $C$ is a constant period.

Guided by the analysis of Kawaler et al.~(2004), we first looked at 
the pulsation frequencies in the December 2009 data. In the December 2009 observations,
a large-amplitude low frequency pulsation was detected at 1097 $\mu$Hz, assumed to be a high order ($n$)
mode similar to the $\Delta n=25$ mode identified by Kawaler et al.~in PG\,1707+427 around
1100 $\mu$Hz.

\begin{figure}
\centerline{\hbox{\psfig{figure=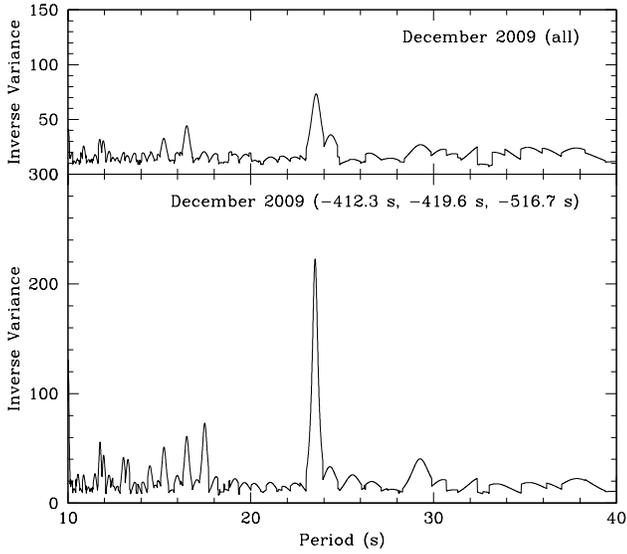,width=8.4cm}}}
 \caption{The inverse variance versus $\Delta P$ for all periods identified in December 2009
(upper panel), and for seven independent periods (all except 412.3, 419.6 and 516.7 s; lower panel).}
 \label{sdss0349fig5}
\end{figure}

To search the suite of periods for a constant period spacing, we used the inverse variance 
method as outlined by O'Donoghue (1994). By letting $\Delta P$ vary between 10 and 40 s (in steps of 0.005 s),
we compared the measured periods with the model periods and plotted the inverse variance of the timing residual (divided 
by $\Delta P$) as a function of $\Delta P$.
The results are shown in the upper panel of Fig.~\ref{sdss0349fig5} for all the 10 periods identified in December 2009.
A low amplitude peak around $\Delta P \sim 23.5$ is evident. The significance of this peak improves substantially
when two periods are removed from the sequence (412.3 and 516.7 s), where the remaining 7 independent periods 
(we have also removed the lower sideband component of the 419-s triplet)
form a well-defined sequence with $\Delta P = 23.52 \pm 0.19$ s and $C = 301.70$ s. This is shown in the bottom panel of
Fig.~\ref{sdss0349fig5}. In comparison, Kawaler et al.~(2004) find $\Delta P = 23.0$ s and $C = 332.9$ s for PG\,1707+427.

\begin{figure}
\centerline{\hbox{\psfig{figure=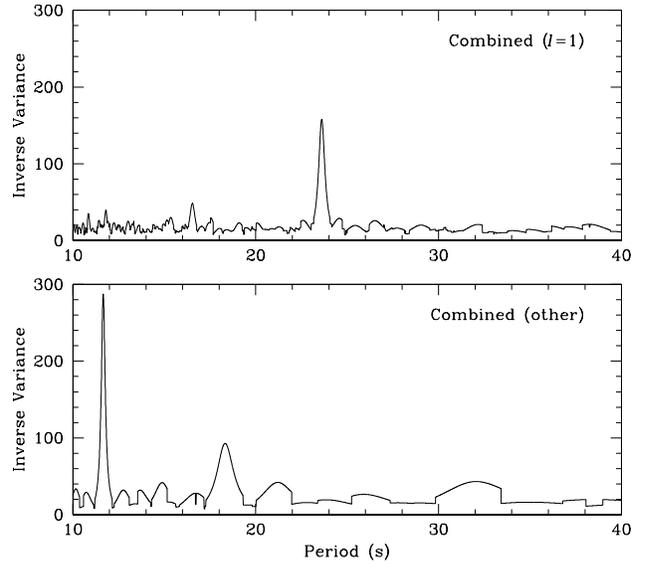,width=8.4cm}}}
 \caption{The inverse variance versus $\Delta P$ for all the periods identified as
$\ell = 1$ modes (upper panel) and for the remaining set of 5 periods that cannot be fitted
by the $\ell = 1$ constant spacing model (lower panel). }
 \label{sdss0349fig6}
\end{figure}

All but five periods identified in 2007 and 2009 fit the asymptotic model with a constant 
period spacing around $\sim$23.5 s. They are listed in Tab.~\ref{sdss0349tab3} against the identified mode
difference ($\Delta n$), the period difference ($\delta P = P_{obs} - P_{model}$) and presumably belong to $\ell = 1$ modes.
The best fit model based on the periods listed in Tab.~\ref{sdss0349tab3} gives
$\Delta P = 23.61 \pm 0.21$ s and $C = 300.54$ s. The upper panel of Fig.~\ref{sdss0349fig6}
shows the inverse variance as a function of $\Delta P$ for this set of periods.
Had we only selected frequencies with amplitudes above
a 5$\sigma$ limit (as opposed to our current 4$\sigma$ selection), the inverse variance analysis would
have given a similar result ($\Delta P = 23.47 \pm 0.13$ s and $C = 301.67$ s).
The sequence of frequencies associated with the best fit model ($\Delta P = 23.61$ s) is shown
in Fig.~\ref{sdss0349fig3} by the lower sequence of short vertical markers (starting at 3327 $\mu$Hz).

\begin{table}
 \centering
 \caption{Periods compared to an asymptotic model with a period spacing of 23.61 s and
a constant period of 300.54 s.}
 \begin{tabular}{@{}ccccrl@{}}
$\Delta n$ & $P_{model}$ & $P_{obs}$ & $\delta P$ & $\delta P$/ & Remarks \\
        & (s)    & (s)    &  (s)   &  23.61 & \\[10pt]
 0   & 300.53 &  300.93  & \hfill   0.40 &\hfill  0.02 & 2007[5.5]/2009[4.3]$^\S$\\
 2   & 347.76 &  349.02  & \hfill   1.26 &\hfill  0.05 & 2007[12.1]   \\
     &        &  349.06  & \hfill   1.30 &\hfill  0.06 & 2009[9.6]\\
 5$^*$ & 418.59& 418.90  & \hfill   0.31 &\hfill  0.01 & 2007[18.6] \\
     &         &  419.18  & \hfill   0.59 &\hfill  0.02 &  2009[15.2]\\
 7   &  465.81 &  465.05  & \hfill --0.76 &\hfill  --0.03 & 2007[3.5] \\
 8   &  489.42 &  486.40$^\dag$  & \hfill --3.02 &\hfill  --0.13 & 2007[4.2] \\
 9   &  513.03 &  511.43  & \hfill --1.60 &\hfill --0.07 &  2009[4.2] \\
11   &  560.25 &  561.83  & \hfill   1.58 &\hfill   0.07 &  2009[4.0] \\
16   &  678.30 &  680.83  & \hfill   2.53 &\hfill   0.11 &  2009[3.6] \\
26   &  914.40 &  911.49  & \hfill --2.91 &\hfill --0.12 &  2009[9.1] \\
     &         &  906.37$^{\ddag}$ & \hfill --8.03 &\hfill --0.34 & 2007[3.9] \\
28   &  961.62 &  963.48  & \hfill   1.86 &\hfill   0.08 & 2007[3.7] \\[5pt]
 \end{tabular}
{\footnotesize
\newline
$^\S$ The year in which the pulsation is detected; the amplitude of
the pulsation in listed in square brackets (in units of mmag).\hfill \\
$^*$ The $m = \pm 1$ modes of this triplet are not listed here. \hfill \\
$^\dag$The one-day alias of the $\Delta n = 8$ mode is at 489.2 s, which gives 
$\delta P$ = --0.2 s and $\delta P$/23.61 = --0.01. \hfill \\
$^\ddag$The one-day alias of the $\Delta n = 26$ mode (2007) is at 916.2 s, which 
gives $\delta P$ = 1.8 s and $\delta P$/23.61 = 0.07. \hfill \\ }
  \label{sdss0349tab3}
\end{table}

The remaining five periods not matched by this model are part of a sequence themselves with
a fixed period spacing of $\Delta P = 11.66 \pm 0.13$ s and $C = 353.91$ s. They are listed
in Tab.~\ref{sdss0349tab4} and the result of the inverse variance test is shown in the lower panel of 
Fig.~\ref{sdss0349fig6}. Within the formal error, this period spacing is consistent with 
being equal to half the period spacing of the $\ell = 1$ model. The sequence of frequencies
associated with this model ($\Delta P = 11.66$ s) is also shown in Fig.~\ref{sdss0349fig3} by the upper
sequence of short vertical markers (starting at 2826 $\mu$Hz) and can be compared with the 
asymptotic model for the $\ell = 1$ modes.

The $\delta P$/$\Delta P$ residuals we obtain for the two models are similar to those
found in studies of other GW Vir stars (e.g., Kawaler et al.~2004).

\begin{table}
 \centering
 \caption{Periods compared to an asymptotic model with a period spacing of 11.66 s and
a constant period of 353.91 s.}
 \begin{tabular}{@{}ccccrl@{}}
$\Delta n$ & $P_{model}$ & $P_{obs}$ & $\delta P$ & $\delta P$/ & Remarks \\
        & (s)    & (s)    &  (s)   &  11.66 & \\[10pt]
 0   & 353.91 &  353.79  & \hfill --0.12 &\hfill --0.01 & 2007[7.2]$^\S$ \\
 5   & 412.21 &  412.27  & \hfill   0.06 &\hfill  0.01  & 2009[10.0]\\
11   & 482.17 &  482.58  & \hfill   0.41 &\hfill  0.04  & 2007[6.6] \\
13   & 505.49 &  504.18 & \hfill --1.31 &\hfill --0.11 & 2007[6.8] \\
14   & 517.15 &  517.84  & \hfill   0.69 &\hfill  0.06  & 2007[11.3] \\
     &        &  516.72  & \hfill --0.43 &\hfill --0.04 & 2009[3.3] \\[5pt]
 \end{tabular}
\newline
\footnotesize{$^\S$ The year in which the pulsation is detected; the amplitude of
the pulsation in listed in square brackets (in units of mmag).\hfill \\
}
  \label{sdss0349tab4}
\end{table}

\section{Discussion}

Our photometric observations show that SDSS\,J0349-0059 is a non-radial pulsator in the 
GW Vir class of variable stars. Frequency splitting of the principal oscillation 
mode at 419 s reveals a rotation period of 0.40 d, if the oscillation mode 
is $\ell = 1$. As with many other GW Vir stars, it is possible to represent most of 
the modes with a linear relationship in period, similar to what has been seen in 
PG\,1707+427, but with parameters that put SDSS J0349 slightly redward of the latter 
star, which now defines the red edge of the GW Vir instability strip.

The five oscillation modes not included in the above group can be fitted by another 
linear relationship in which the spacing is a harmonic of the first sequence. However, 
there is no currently known physical model which explains this behaviour. This is unlike
PG\,1707+427 where the discrepant oscillation modes appear to have an $\ell = 2$ origin.

\section*{Acknowledgments}

Our research is supported by the University of Cape Town and by 
the National Research Foundation. This work was completed during a
visit by PAW to the University of Southampton, which was funded by 
an ERC advanced investigator grant awarded to R.~Fender.
This paper uses observations made at the South African Astronomical 
Observatory (SAAO).


\begin{thebibliography}{99}
\bibitem{aih11}   Aihara H., et al., 2011, ApJS, 193, 29
\bibitem{cor06}   C\'orsico A.H., et al., 2006, A\&{A}, 458, 259
\bibitem{hug06}   H\"ugelmeyer S.D., et al., 2006, A\&{A}, 454, 617
\bibitem{kaw04}   Kawaler S.D., et al., 2004, A\&{A}, 428, 969
\bibitem{dod94}   O'Donoghue D., 1994, MNRAS, 270, 222
\bibitem{dod95}   O'Donoghue D., 1995, Baltic Ast, 4, 517
\bibitem{qui09}   Quiron P.-O., 2009, Comm. in Astroseismology, 159, 99
\bibitem{qui07}   Quiron P.-O., Fontaine G., Brassard P., 2007, ApJS, 171, 219
\bibitem{win08}   Winget D.E., Kepler S.O., 2008, ARA\&{A}, 46, 147
\end{thebibliography}
\end{document}